\hoffset = -10mm

\vglue1cm \line {\vbox{\halign{\hfill#\hfill\cr Nice INLN 95/25 \cr}}
\hfill\vbox{\halign{\hfill#\hfill\cr April 1996\cr}}} \vglue3cm

\centerline { {\bf ON THE LIGHT CONE SINGULARITY }} \smallskip
\centerline {{\bf OF}} \smallskip \centerline {{\bf THE THERMAL
EFFECTIVE EXPANSION}} \bigskip \bigskip\bigskip\bigskip\bigskip
\centerline { Thierry Grandou } \bigskip\medskip

\bigskip \centerline {  Institut Non Lin\'eaire de Nice UMR CNRS 129}

\centerline {1361, route des lucioles, 06560 Valbonne, France}
\centerline {e-mail:grandou@doublon.unice.fr}
\bigskip\bigskip\medskip\bigskip\bigskip\bigskip \centerline{\bf
ABSTRACT} \bigskip\smallskip\noindent We consider a scalar massless
quantum field model, at finite temperature $T$, both renormalizable and
asymptotically free. Focussing on the singular structure of the
effective perturbation theory about the light cone, several new insights
are put forth, regarding the interplay between hard thermal loop
resummation and the overall compensation of collinear singularities.

\vfill  \eject

{\bf 1. Introduction} \bigskip

The inherent non perturbative character of finite temperature quantum
field theories has been recognized a few years ago on the basis of
general group theoretical arguments [1]. At the same time, the formal
perturbative series themselves were shown to necessitate a full
redefinition of their original form so as to yield (hopefully) sensical
results. This is the resummation program devised by Braaten and Pisarski
and also by Frenkel and Taylor [2], hereafter referred to as "effective
theory" for short. \medskip\noindent In order not to be just an academic
recipe, such a program implicitly requires that the original and/or
re-organised perturbative series be infrared safe. In the recent years,
a great deal of efforts has been devoted to the study of the infrared
structure of the original perturbative series, with the conclusion that,
roughly speaking, the situation at non zero temperature ($T\neq 0$) was
not worse, globally, than at zero temperature ($T=0$) [3].
\medskip\noindent But re-defining a perturbative series is rarely
trivial an operation. Properties which were known to hold true for the
original series may become much harder to control in the redefined one.
In our opinion, the collinear singularity recently found in thermal QCD
by using the effective perturbative series [4,5] might reveal a salient
illustration of this fact. Note that we are aware that this point of
view differs, at first sight, from the ones adopted in the solutions
proposed in [6] and [7]. Very recently, though, it has been realized
[18], that some unexpected connection could exist with the analysis
proposed in [6]. In this paper we will restrict ourselves to a scalar
model and analyse the singular structure of the effective theory in much
details. \bigskip\noindent A preliminary report of the present work
appeared in [8], which is here corrected, completed and enlarged. The
paper is organized as follows. Section 2 is an introduction to the
model, and the necessary elements of the particular real time formalism
which is used, are given. In section 3, the singularity structure of a
whole series of diagrams is investigated, deferring some lengthy details
to an appendix. The results of this section rely on an approximation
which consists in keeping only the potentially most singular part of the
hard thermal loop (HTL) self energy. This approximation gets completed
in section 4, where the other terms are taken into account. Further
consequences are drawn in sections 5 and 6, respectively related to the
problem of the $T=0$ and $T\neq 0$ singularity mixing, and to the
possibility of an unambiguous renormalisation constant definition in a
thermal context. Eventually, a conclusion gathers our
results.\bigskip\bigskip\bigskip {\bf 2. The model} \bigskip In order to
get rid of unessential complications, we deal with the toy model
provided by a massless hermitian scalar quantum field. It is endowed
with a cubic self interaction, of coupling strength $g$, renormalisable
at $D=6$ space-time dimensions. Furthermore, its quanta are assumed to
form a plasma in thermodynamical equilibrium at temperature $T$.
Infrared as well as ultraviolet singularities are taken care of by
working at $D=6\pm 2\varepsilon$ space-time dimensions respectively.
Also, we will use the convention of upper case letters for $D$-momenta
and lower case ones for their components, i.e, $P=(p_0,{\vec
p})$.\medskip\noindent Use will be made of the material and results of a
previous work [9] where a calculation performed at second non trivial
order of bare perturbation theory, $O(g^4)$, was able to display the
overall compensation of infrared and collinear singularities for some
particular topology. \medskip\noindent The process under consideration
consists in a "Higgs" particle of $D$-momentum $Q=(q,{\vec {0}})$ in the
rest-frame of a plasma in equilibrium at temperature $T$. But, contrary
to the case studied in [9], this $D$-momentum is here assumed to be
soft, that is $q=O(g T)$. Also, our analysis will take place in the
framework of the real time formalism, within "the basis" provided by the
retarded and advanced free field functions, $\Delta_R (K)$ and $\Delta_A
(K)$ respectively [5]. At any step, though, we have checked that the
same expressions are obtained in the more customary real time formalism
of the Kobes and Semenoff rules (the "1/2" formalism,
say).\medskip\noindent The real part of the thermal one loop self energy
reads [9] $${\rm Re}\Sigma_T(P)={2g^2 \over
{(4\pi)^{3+\varepsilon}\Gamma(2+\varepsilon)} }\int_0^{\infty} {\rm d}k\
k^{D-3}n(k) \sum_{\eta=\pm 1} {\bf P}\int_{-1}^{1}{\rm d}x {
(1-x^2)^{1+\varepsilon}\over {P^2+2k(\eta p_0-px)} }
\eqno(2.1)$$\noindent where a combinational weight factor $1/2$ is taken
into account, and where ${\bf P}$ stands for a principal value
prescription. Also, throughout the calculations, use will be made of the
following relations for statistical factors $$n(k_0) = (e^{\beta k_0} -
1)^{-1},\ n^B(k_0) = (e^{\beta \vert k_0\vert} - 1)^{-1},\
\varepsilon(k_0)(1+2n(k_0))=1+2n^B(k_0) \eqno(2.2)$$ The phase-space
domain where collinear singularities come from has been recognized to be
determined by the condition $P^2<<p^2<<T^2$. In this domain, the
function $\{Re\Sigma_T\}$, that we choose to describe in terms of the
two independent variables $P^2$ and $p^2$, is approximated by a real
function $A$, defined for all $P^2$ values $$A(P^2,p^2)=-m^2\ {1\over
\varepsilon}\ {P^2\over p^2}\Biggl\lbrace 1-{1+2^{1+
\varepsilon}\varepsilon\over 4^{\varepsilon}}\ {\vert{P^2\over
p^2}\vert}^\varepsilon \left (\Theta(P^2)+\cos
(\pi\varepsilon)\Theta(-P^2) \right)\Biggr\rbrace \eqno(2.3)$$ where
$m^2$, the thermal mass squared, is given by $$m^2=2 {g^2\over
(4\pi)^3}\zeta(2+2\varepsilon)T^{2+2\varepsilon}\eqno(2.4)$$ Expression
(2.3) displays the full HTL (that is, leading, order $g^2T^2$ part)
entailed in the real part of the thermal self energy function. Only for
space-like values of its argument $P^2$, does the thermal self energy
function develop an imaginary part of same leading order $g^2T^2$. This
corresponds to the so called "Landau damping mechanism", and one can
write eventually $$\Sigma_F=A+iB\ ,\ \ \ \ \ \ B(P^2,p^2)=-\pi
m^2(-{P^2\over p^2})^{1+\varepsilon}\Theta(-P^2){1+2^{1+
\varepsilon}\varepsilon\over 4^{\varepsilon}}\
{\sin(\pi\varepsilon)\over \pi\varepsilon } \eqno(2.5)$$ where
$\Sigma_F$ denotes the Feynman self energy. Indeed, the functions $A$
and $B$ can be seen as obtained through the usual Feynman prescription
$\{p_0\longrightarrow p_0+i\eta p_0\}$, out of $one$ analytical self
energy function $\Sigma$ defined in the complex energy plane
$$A=\lim_{\eta=0}\ {1\over 2}\left(\Sigma(p_0+i\eta
p_0,p)+\Sigma(p_0-i\eta p_0,p)\right)$$ $$B=\lim_{\eta=0}\ {1\over
2i}\left(\Sigma(p_0+i\eta p_0,p)-\Sigma(p_0-i\eta p_0,p)\right)$$with
$\Sigma$ the light cone approximate self energy $$\Sigma=-m^2\ {1\over
\varepsilon}\ {P^2\over p^2}\Biggl\lbrace 1-{1+2^{1+
\varepsilon}\varepsilon\over 4^{\varepsilon}}\left({P^2\over
p^2}\right)^\varepsilon \Biggr\rbrace \eqno(2.6)$$ Note that in (2.6),
the limit ${\varepsilon =0}$ can be safely taken and that the thermal
self energy function is a regular function of $P^2$ in any neighbourhood
of the light cone. One gets effectively $$\Sigma(P^2,p^2)=-2m^2{P^2\over
p^2}\left({1\over 2}\ln({4p^2\over P^2})-1 \right)\eqno(2.7)$$which,
expressed in terms of the variables $P^2$ and $p^2$, is nothing but the
more familiar form [12] of the HTL self energy of the model, with its
second kind Legendre function $Q_1(p_0,p)$ approximate form at
$|p_0|\sim p$. In the $R/A$ real time formalism we will be using,
retarded and advanced functions must be used. They are simply related to
the Feynman ones by the set of relations
$$\Sigma_R(P)=\Theta(p_0)\Sigma_F(P)+\Theta(-p_0)\Sigma_F^*(P),\ \ \
\Sigma_A(P)=\Theta(p_0)\Sigma_F^*(P)+\Theta(-p_0)\Sigma_F(P)\eqno(2.8)$$
In the region ${\cal D}$ where one has $P^2<<p^2<<T^2$ and ${ O}
(p_\mu)=gT$, effective propagators ${}^*\Delta_\alpha (P)$ must be used
so as to get the full leading order $g^2T^2$ correction to the free
quantities, with $${}^*\Delta_\alpha (P)={i\over
{P^2-\Sigma_\alpha(P^2,p^2)+i\varepsilon_\alpha p_0}}\ ,\ \ \ \
\alpha=R,A.\eqno(2.9)$$ A point to be stressed is that the domain which
requires that effective perturbation theory be used, is also the domain
where mass (or collinear) singularities stem from, thereby opening up
the very "window" where the light cone singular behaviour of the
effective thermal expansion must be studied. \medskip\smallskip\noindent
An important remark is in order. Admittedly, there is no HTL in the
three-point function of the $(g\varphi^3)_6$ theory [10], in the sense
that $g^2T^2$ leading parts, eventually, cancel out, leaving order
$g^2T^{3/2}$ pieces at most. However, in [10], not all the terms were
considered and, more generally, it should be noted that this statement
is not sufficient for ignoring the one loop three point function. For
example, it has been stressed recently [7] that, over collinear
configuration of external soft momenta, a special enhancement of bare
orders of magnitude, results out of collinear singularities, leading to
a breakdown of the HTL resummation program. However, besides the new
internal difficulties encountered by the improved HTL resummation
program itself [24], it is important to realize that such a situation
does not show up here. This is basically because the Higgs particle has
timelike $D$-momentum $Q=(q,{\vec 0})$. Explicit calculation then shows
that no collinear singularity ever develop in that case. Now, a wider
analysis of the three point function is certainly worth undertaking, but
falls beyond the scope of the present article where one loop vertex
corrections will not be considered.\bigskip\bigskip\bigskip

{\bf 3- Singular structure} \bigskip We begin with considering the case
of $N$ one-loop self energy insertions along the $P$-line as depicted on
Fig.1, with $N'=0$. Then, the retarded "polarisation tensor"
$\Pi^{(N)}_{RR}(Q)$ of the"Higgs particle" (hereafter written
$\Pi^{(N)}_{R}(Q)$ for short) can be shown to admit the following
expression (see (61) in [5]) $$\displaylines{\Pi^{(N)}_R(Q)=-i\int {{\rm
d}^DP\over (2\pi)^D}(1+2n(p_0))\ {\rm Disc}_P\left(\Delta_R^{(N)}
(P)\Delta_R (P')V^2_{RRA}(P,Q,-P')\right)\qquad(3.1)}$$ where $V_{RRA}$
is the bare vertex with two external lines of the retarded type,
corresponding to $D$-momenta $P$ and $Q$, and one advanced external line
of $D$-momentum $-P'$. The convention for the flow of momenta is that
their sum vanishes: $P+Q+(-P')=0$. At zeroth order, $V_{RRA}$ is simply
given by the "electric charge", $e$, which couples the Higgs particle to
the hermitian scalar field. Eventually, the prescription "${\rm
Disc_P}$" means that the discontinuity in the variable $p_0$ is to be
taken. In the $\{R/A\}$ real time formalism we are using, this is simply
achieved by writing $${\rm
Disc_P}F_{R\beta\delta}(P,Q,R)=F_{R\beta\delta}(P,Q,R)-F_{A\beta\delta}(P,Q,R)
\eqno(3.2)$$ with $\beta$ and $\delta$ any of the two $R,A$
possibilities. \smallskip \smallskip\noindent In above equation (3.1),
$\Delta_R^{(N)}$ denotes a $N$ one-loop self energy corrected free field
function, and $\Pi^{(N)}_R(Q)$, the ensuing corrected tensor. Omitting
unessential factors, its imaginary part reads $$\displaylines{ {\rm
Im}\Pi^{(N)}_R(Q)=\int {\rm
d}^DP(1+2n(p_0))\varepsilon(p'_0)\delta(P'^2)\{{(-1)^N\over
N!}\pi\varepsilon(p_0)\delta^{(N)}(P^2){\rm Re}(\Sigma_R^N(P))-{\bf
P}{{\rm Im}(\Sigma_R^N(P))\over (P^2)^{N+1}}\}\qquad(3.3)}$$ where
$\varepsilon(p_0)$ is the sign of $p_0$, and $\delta^{(N)}$ the $N^{th}$
derivative of the $\delta $-distribution. In view of equations (2.3),
(2.5) and of their bounded $P$-domain, note that the real and imaginary
parts of the $\{\Sigma_R^{HTL}(P)\}^N$-insertions are acceptable
test-functions, yielding well defined expressions when acted upon with
$\delta^{(N)}(P^2)$ and ${\bf P}/(P^2)^{N+1}$ distributions
respectively. How things are working out is instructive to observe. At
$N=1$, we have from the $\delta^{(1)}(P^2)$-term
$$\displaylines{{-e^2\over (2\pi)^{D-1}}\int{\rm d}^{D-1}p\int{{\rm
d}P^2\over 2p_0}(1+2n(p_0))\varepsilon(p_0)\cr\hfill\times\
\varepsilon(p'_0)\delta(P'^2)\pi\delta(P^2) {{\rm d}\over {\rm d}
P^2}A(P^2,p^2)\qquad(3.4)}$$ where $|p_0(P^2,p^2)|={\sqrt {P^2+p^2}}$.
Substituing (2.3), we get immediately $$-{\pi e^2 m^2\over
(2\pi)^{D-1}}{\Omega_D\over 4q}({q\over 2})(1+2n^B({q\over 2}))({1\over
\varepsilon})\eqno(3.5)$$ where $\Omega_D$ is the total solid angle of
the model. The singular piece coming from the term involving the
principal value, more involved, is $$\displaylines{-{e^2\over
(2\pi)^{D-1}}{1+2^{1+ \varepsilon}\varepsilon\over 4^{\varepsilon}}\
{\sin(\pi\varepsilon)\over \pi\varepsilon }\int{{\rm d}^{D-1}p\over
2p}(1+2n(p))\varepsilon(p'_0)\delta(P'^2)\cr\hfill \times\ \ {\bf
P}\int{{\rm d}P^2\over (P^2)^2}\left(-\pi m^2(-{P^2\over
p^2})^{1+\varepsilon}\right)\Theta(-P^2)\qquad(3.6)}$$ where (2.5) has
been used. By integrating over $P^2$ first, which yields a
$\varepsilon^{-1}$ mass singularity, and then over $p$, one gets $$+{\pi
e^2m^2\over (2\pi)^{D-1}}{\Omega_D\over 4q}({q\over 2})(1+2n^B({q\over
2}))({1\over \varepsilon})\ {1+2^{1+ \varepsilon}\varepsilon\over
4^{\varepsilon}}\ {\sin(\pi\varepsilon)\over \pi\varepsilon
}\eqno(3.7)$$ The sum of the two terms displays the compensation of the
collinear singularity, as noticed long ago by many authors for scalar as
well as for gauge theories.\medskip\noindent At $N=2$, one has ${\rm
Re}\Sigma^2_R(P)=A^2(P)-B^2(P)$. The function $B$ is of the order of
$(-P^2)^{1+\varepsilon}$ with $\varepsilon$ positive, and therefore does
not contribute when acted upon with the $\delta^{(2)}(P^2)$
distribution. Indeed, for exactly the same reason, only the
$\{A^N\}$-part of ${\rm Re} \Sigma^N_R(P)$ will ever contribute at
general $N$, because of the $\delta^{(N)}(P^2)$ distribution. In the
present case, $N=2$, one gets $$-\pi{e^2\over
(2\pi)^{D-1}}{\Omega_D\over 4q} {2\over q}\left(1+2n^B({q\over
2})\right)({m^2\over \varepsilon})^2\eqno(3.8)$$ which, by recalling
that $q$ is of the order of $gT$, is consistently seen to be on the same
order as (3.5) and (3.7) at $N=1$. Now, the part involving the principal
value distribution reads $$\displaylines{ -\pi{e^2\over
(2\pi)^{D-1}}{1+2^{1+ \varepsilon}\varepsilon\over
4^{\varepsilon}}{\sin(\pi\varepsilon)\over \pi\varepsilon }\int{\rm
d}^{D-1}p \int{{\rm d}P^2\over 2p_0}
(1+2n(p_0))\varepsilon(p'_0)\delta(P'^2)\left(-{{\bf P}\over
(P^2)^3}\right)\hfill\cr \hfill \times\
2\varepsilon(p_0)\left(-{m^2P^2\over \varepsilon p^2}\ (1-{1+2^{1+
\varepsilon}\varepsilon\over 4^{\varepsilon}}\
{\sin(\pi\varepsilon)\over \pi\varepsilon }\vert {P^2\over
p^2}\vert^\varepsilon \cos(\pi\varepsilon))\right)\left(-\pi
m^2(-{P^2\over p^2})^{1+\varepsilon}\right)\Theta(-P^2)\qquad(3.9)
\cr}$$ Integrating out over $P^2$ and $p$, one gets $$+\pi{e^2\over
(2\pi)^{D-1}}{\Omega_D\over 4q} {2\over q}\left(1+2n^B({q\over
2})\right){1+2^{1+ \varepsilon}\varepsilon\over 4^{\varepsilon}}\
{\sin(\pi\varepsilon)\over \pi\varepsilon }(-{m^4\over
\varepsilon})\left({1\over \varepsilon}-{1+2^{1+
\varepsilon}\varepsilon\over 4^{\varepsilon}2\varepsilon
}\cos(\pi\varepsilon)\right)\eqno(3.10)$$ Expanding in powers of
$\varepsilon$ the terms $${1+2^{1+ \varepsilon}\varepsilon\over
4^{\varepsilon}}\ {\sin(\pi\varepsilon)\over \pi\varepsilon
}\left({1\over \varepsilon}-{1+2^{1+ \varepsilon}\varepsilon\over
4^{\varepsilon}2\varepsilon }\cos(\pi\varepsilon)\right)={1\over
2\varepsilon}\ \left(1+O(\varepsilon^2)\right)\eqno(3.11)$$ the sum of
the two terms (3.9) and (3.10) is readily seen to display the expected
singularity compensation. Indeed, inspection of the first values of $N$
($1\leq N\leq 5$), shows that one may drop the terms
$2^\varepsilon,4^{-\varepsilon}$ appearing in (2.3) and (2.5) without
spoiling at all the singularity cancellation mechanism. In what follows,
we will take advantage of that simplification in order to avoid
overcharging with details.\medskip\noindent At $N=3$, one has
$$\Sigma_R^3(P^2,p^2)=(A^3-3AB^2)(P^2,p^2)\ +\
i\varepsilon(p_0)(3A^2B-B^3)(P^2,p^2) \eqno(3.12)$$ Of the real part,
only the $\{A^3\}$ term will contribute according to the remark above.
The resulting singularity, the stronger, of order $\varepsilon^{-3}$ is
given by $$\left((-1)^3/ 3!\right)\pi\varepsilon(p_0)\delta^{(3)}(P^2)\
A^3(P^2,p^2)=\pi \varepsilon(p_0) \delta(P^2)\ \left(-{m^2/ \varepsilon
p^2}\right)^3\eqno(3.13)$$ where, for short, we do not write the
remaining integration over $p$ (see the remark after (3.20)). The
imaginary term $\{-B^3\}$ is easily integrated over $P^2$ and yields
$$\varepsilon(p_0)\ (1+2\varepsilon)^3\ ({\sin(\pi\varepsilon)\over
\pi\varepsilon })^3\ (-{\pi m^2\over p^2})^3\ ({1\over
3\varepsilon})\eqno(3.14)$$ The other term $\{3A^2B\}$ is a bit more
involved. Integrated over $P^2$ it results in $$\varepsilon(p_0)\ 3({\pi
m^2\over p^2})(-{m^2\over\varepsilon p^2})^2\ (1+2\varepsilon)\
({\sin(\pi\varepsilon)\over \pi\varepsilon })\Biggl\lbrace {1\over
\varepsilon}-{2(1+2\varepsilon)\cos(\pi\varepsilon)\over
2\varepsilon}+{(1+2\varepsilon)^2\cos^2(\pi\varepsilon)\over
3\varepsilon}\Biggr\rbrace \eqno(3.15)$$ Expanding in powers of
$\varepsilon$ the two trigonometric functions, we get
$$\varepsilon(p_0)3({\pi m^2\over p^2})(-{m^2\over\varepsilon p^2})^2\
{1\over 3\varepsilon}\left(1+{(\pi\varepsilon)^2\over
3}+O(\varepsilon^3)\right)\eqno(3.16)$$ The mass singularity
compensation is now made obvious as the first term in the last
parenthesis of (3.16), the $+1$, compensates for the real term (3.13),
whereas the second term and (3.14) cancel each other, just leaving
regular contributions. Note that with respect to the previous case,
$N=2$, expanding the trigonometric functions has become mandatory in
order to manifest the singularity compensation.\smallskip\noindent As we
have been able to check that a similar compensation of all leading and
subleading mass singularities, occur at $N=4$ and $N=5$ as well, these
results strongly suggest that, for all $N$, such an overall cancellation
of mass singularities is extremely likely to take place, and not just at
$N\leq 2$, as was incorrectly stated in [8]. However, the control of the
ensuing finite parts and their subsequent summation over $N$, remains
somewhat puzzling. This is why we will carry out the summation over $N$
a different way.\bigskip\noindent First, we take the complicated
expressions (2.3) and (2.5) to the much simpler forms $${
A}(P^2,p^2)=-m^2\ {1\over \varepsilon}\ {P^2\over p^2}\bigl\lbrace
1-{\vert{P^2\over p^2}\vert}^\varepsilon \bigr\rbrace\ ,\ \ \ \ {
B}(P^2,p^2)=-\pi m^2(-{P^2\over
p^2})^{1+\varepsilon}\Theta(-P^2)\eqno(3.17)$$ Of course, with (3.17) as
a crude approximation of (2.3) and (2.5), only the leading mass
singularities cancel out for all $N$, while leaving a finite series of
subleading mass singularities. The latter, though, can be resummed into
regular functions. Then, bringing the expressions (3.17) to completion,
that is to the forms (2.3)-(2.5), the full result can be derived rather
easily out of those regular functions. This procedure will further open
two interesting possibilities, dealt with in sections 5 and 6. Also, it
applies straightforwardly to the (fermionic) cases where HTL self
energies are proportional to the second kind Legendre function
$Q_0(p_0,p)$ rather than $Q_1$. \medskip\noindent We begin with showing
that for all $N$, the stronger singularities, of order
$\varepsilon^{-N}$ cancel each other. In effect, at general $N$, such a
singularity comes from the real part of $\Sigma_R^N$. It is
$$\left((-1)^N/ N!\right)\pi\varepsilon(p_0)\ \delta^{(N)}(P^2)\ {
A}^N(P^2,p^2)=\pi \varepsilon(p_0) \delta(P^2)\ \left(-{m^2/ \varepsilon
p^2}\right)^N\eqno(3.18)$$ whereas that very contribution from the
imaginary part of $\Sigma_R^N$ can be written $$ -\varepsilon(p_0)\pi
m^2(-{m^2\over \varepsilon p^2})^{N-1} C^1_N \int_{-p^2}^0 {{\rm
d}P^2\over (P^2)^{N+1}}\left(-{P^2\over p^2}\right)^{1+\varepsilon}
(P^2)^{N-1}(1-\vert {P^2\over p^2}\vert^\varepsilon)^{N-1}\eqno(3.19)$$
where the $C^k_N$ are the binomial coefficients, and where the
$\Theta(-P^2)$ distribution has been turned into the
$[-p^2,0]$-integration range, by taking the condition $P^2<<p^2<<T^2$
into account. The integral over $P^2$ splits into $N$ integrals, the sum
of which reads [11] $$(p^{-2})\left({1\over \varepsilon}\right)\ \left(
C^1_N\ \sum_{k=0}^{N-1}\ {C_{N-1}^k(-1)^k\over k+1}\ =\ 1\right)
\eqno(3.20)$$ By putting this expression back into (3.19), one gets the
exact compensation of the $\varepsilon^{-N}$ singularity of equation
(3.18), thus completing the proof that the two most singular terms
cancel each other at any number $ N$ of one-loop self energy insertions
along the $P$-line. \smallskip\noindent It is interesting to note that,
for this process at least, the choice of the two independent variables
$P^2$ and $p^2$ is a most convenient one. In particular, singularity
compensation is rendered manifest after a unique one dimensional
integration, the one over $P^2$, is performed. Also, it displays the
mass (or collinear) singularity character of the encountered poles. Had
we choosen $p_0$ and $\vert {\vec p}\vert$ as another, more customary
choice of two independent variables, none of these two properties were
obtained. \bigskip\noindent However a milder singularity (like (3.14))
remains non cancelled. The phenomenon of course starts out at $N=3$ with
the $\{B^3\}$ term, and is, thereof, general. For any $N=2k+1$ with
$k\geq 1$, the imaginary part of $\Sigma_R^N$ develops a similar
singularity under integration over $P^2$
$$(i\varepsilon(p_0)B)^{2k+1}\Rightarrow
i\varepsilon(p_0)(-1)^k{\Theta(-P^2){\bf P}\over
(-P^2)^{1-(2k+1)\varepsilon}} (\pi {m^2\over
p^{2+2\varepsilon}})^{2k+1}\Rightarrow {i\varepsilon(p_0)(-1)^k\over
(2k+1)\varepsilon}({\pi m^2\over p^{2+2\varepsilon}})^{2k+1}\eqno
(3.21)$$ In the imaginary part of $\Sigma_R^N$, but for $N$-values
greater than $2k+1$ at some given value of $k\geq 1$, each of these
terms do appear again, multiplied by some powers of $A(P^2,p^2)$.
Typically, $$\varepsilon(p_0)(-1)^kC^{2k+1}_NA^{N-2k-1}B^{2k+1},\ \ \ \
N\geq 2k+1,\ \ \ \ k\geq 1 \eqno(3.22)$$ Selecting large enough a value
of $N$, the sum of these terms from ${\cal N}=2k+1$ to ${\cal N}=N$,
say, reads $$\varepsilon(p_0)(-1)^k(-\pi {m^2\over p^{2+2\varepsilon}}
)^{2k+1}{\bf \sum}_{{\cal N}}(-{m^2\over \varepsilon p^2})^{{\cal
N}-2k-1} C_{{\cal N}}^{2k+1}\int_{-p^2}^0 {\rm d}P^2{\bf P}{(1-\vert
{P^2/ p^2}\vert^\varepsilon )^{{\cal N}-2k-1}\over
(-P^2)^{1-(2k+1)\varepsilon}}\eqno(3.23)$$ Carrying out the integration
over $P^2$ and taking advantage of the arithmetical identity (see
Appendix) $$C_{{\cal N}}^{2k+1}(\sum_{j=0}^{{\cal N}-2k-1}\ {(-1)^j\over
j+2k+1} C_{{\cal N}-2k-1}^j)={1\over 2k+1}\eqno(3.24)$$ the expression
(3.23) can be given the much simpler form $$\varepsilon(p_0)(-1)^k(-\pi
{m^2\over p^{2+2\varepsilon}})^{2k+1}\ {1\over (2k+1)\ \varepsilon}\ \
{\bf \sum}_{{\cal N}=2k+1}^{{\cal N}=N}\ (-{m^2\over \varepsilon
p^2})^{{\cal N}-2k-1}\eqno(3.25)$$ Summing over $N$ the imaginary part
of $\Pi^{(N)}_R$, and letting $N$ tend to infinity, one can re-arrange
the sum of the series by singling out the $\{B^{2k+1}\}$-terms we have
just analysed. One gets $$\varepsilon(p_0)(-1)^k\ (-\pi {m^2\over
p^{2+2\varepsilon}})^{2k+1}{1\over (2k+1)\ \varepsilon}\ \times \
Z(\varepsilon,p)\ ,\ \ \ Z(\varepsilon,p)=\left(1-(-{m^2\over
\varepsilon p^2})\right)^{-1}\eqno(3.26)$$ The occurence of the function
$Z$ is quite remarkable as it can be obtained out of the real part of
the self energy by differentiating it with respect to $P^2$, at $P^2=0$.
We have, the prime indicating such a derivation
$$Z(\varepsilon,p)=\left(1-{\rm
Re}\Sigma'(0,p^2)\right)^{-1}=\left(1+{m^2 / \varepsilon
p^2}\right)^{-1}\eqno(3.27)$$ Eventually, summing over $k$, we get
$$\sum_N^\infty\int {\rm d}P^2{{\cal{\bf P}}\over
(P^2)^{N+1}}\sum_{k=1}^{2k=N-1-\Theta((-1)^N)}
C_N^{2k+1}(-1)^kB^{2k+1}A^{N-2k-1} =\varepsilon(p_0)\
{Z(\varepsilon,p)\over \varepsilon}F(p)\eqno(3.28)$$ where one has
$$\lim_{\varepsilon =0}\ {Z(\varepsilon,p)/ \varepsilon}={p^2/ m^2},\ \
\ \ F(p)=-{1\over 12}\left({g\over 4}( {T\over
p})^{1+\varepsilon}\right)^2+{\rm tg}^{-1} \Biggl\lbrace{1\over
12}\left({g\over 4}({T\over p})^{1+\varepsilon}\right)^2\Biggr\rbrace
\eqno(3.29)$$ two regular functions of $p$ in the domain ${\cal D}$.
Note that for the series (3.28) to converge and yield the function
$F(p)$, it is necessary that the condition $$1\geq {1\over 12}({gT\over
4p})^2 \ \Longrightarrow \ p\geq ({1\over 8{\sqrt 3}})gT\eqno(3.30)$$ be
fulfilled. At its turn, (3.30) is obviously consistent with the assumed
softness of the $P$-line. The convergence of the series displayed in
(3.25) and leading to (3.27) is less obvious. This is because
$m^2/\varepsilon p^2$ cannot be expected to lie within the convergence
radius of (3.25) when $\varepsilon^{-1}$ regularizes a potentially
divergent behaviour. As should be made clear through sections 5 and 6,
though, (3.27) can be provided with the same rigorous proof as the one
developed at $T=0$, relying on the renormalization group equations and
the asymptotic freedom property of the model [13]. \bigskip\noindent To
summarize, we have obtained $${\rm Im}\Pi_R(Q)={e^2\over
(2\pi)^{D-1}}({q\over 2})^{D-3}{\Omega_D\over 4q}(1+2n^B({q\over 2}))
\left(Z(\varepsilon,q/2)/ \varepsilon\right)F({q/ 2})\eqno(3.31)$$ an
infrared safe result. Up to the Born term which corresponds to zero self
energy insertion, $N=0$, the result (3.31) should coincide, by
construction, to a calculation of ${\rm Im}\Pi_R(Q)$ in which a bare
internal $P$-line has been replaced by the corresponding effective
propagator ${}^*\Delta_R (P)$. When $q$ is soft, though, this only
replacement is not able to yield the full order $g^2T^2$ correction to
the zeroth order calculation, and a similar replacement must be
envisaged for the second internal line as well. \bigskip\noindent Within
obvious notations, the imaginary part of the Higgs "polarisation tensor"
obtained by inserting $N$ and $N'$ self energy corrections along the $P$
and $P'$ lines, see Fig.1, can be given the compact form $${\rm Im}
\Pi^{(N,N')}_R(Q)={2e^2\over (2\pi)^D}\int {\rm
d}^DP(1+2n(p_0))\bigl\lbrace {\rm Im}(\Delta^{(N')}_R(P')){\rm
Disc_P}(\Delta^{(N)}_R(P))+(N\leftrightarrow N')\bigr\rbrace
\eqno(3.32)$$ where ${\rm Im}\Delta^{(N')}_R(P')$ can be read from
equation (3.3) right hand side by replacing $P$ and $N$ by $P'$ and
$N'$. Writing the contribution to ${\rm Im}\Pi^{(N,N')}_R(Q)$ of a
generic term of ${\rm Im}\Sigma^{(N')}_R(P')$ (see (3.3)), one gets a
rather cumbersome expression which is given in the appendix. The key
observation, though, is as follows. Both $P$ and $P'$ lines develop
singular behaviours on the light cone, but for the $P'$ line, the
condition $ P'^2=0$ gets translated into the condition $\{2p-q=0\}$, and
the corresponding light cone singular behaviours into generic factors
$$\Theta(2p-q)[q(2p-q)]^{-1+(2k'+1)\varepsilon}\eqno(3.33)$$The point is
that integrations over $P^2$ and then over $p$, just decouple and get
the two mass singular behaviours factorised. This amazing simplification
is entirely due to the peculiarity of the process under consideration,
with ${\vec q}= 0$, and does not seem to extend beyond. Having
integrated over $P^2$, and summed over $N$, one gets effectively $$
\sum_N^\infty {\rm Im} \Pi^{(N,N')}_R(Q)=\int {{\rm d}^{D-1}p\over
2p}(1+2n^B(p)){\rm Im}(\Delta^{(N')}_R(q-p,{\vec p}) )\
({Z(\varepsilon,p)/ \varepsilon})F({p})\eqno(3.34)$$ Now the functions
$Z/\varepsilon$ and $F$ of (3.29) are perfectly regular at $p=q/2$, so
that the same pattern as before, at $N'=0$, applies here again for
$\Delta^{(N')}_R(P')$. Using the $P'$-adapted form of identity (3.20),
that is, explicitly $$C^1_{N'}\int_{q/2}^q{{\rm d}p\over
[q(2p-q)]^{1-\varepsilon}}\left(1-\vert {q(2p-q)\over
4p^2}\vert^\varepsilon\right)^{N'-1}={q^{2\varepsilon}\over 2q}{1\over
\varepsilon}\eqno(3.35)$$ it is shown in the appendix that the two most
singular terms, of order $\varepsilon^{-N'}$ compensate each other,
while leaving a finite series of sub-leading mass singularities. Summing
the latter series, the final result reads $${\rm Im}{}\Pi_R(q,{\vec
0})={e^2\over (2\pi)^{D-1}}({q\over 2})^{D-3}{\Omega_D\over
4q}(1+2n^B({q\over 2})) ({Z(\varepsilon,q/2)/ \varepsilon})^2 F^2({q/
2}) \eqno(3.36)$$ It is infrared stable and in line with (3.31) on view
of the peculiar $\{P,P'\}$ internal line symmetry. By construction,
summing over $N$ and $N'$ from zero to infinity, just replaces both
internal lines by their corresponding effective propagators,
${}^*\Delta_R(P)$ and ${}^*\Delta_R(P')$, so that the above result
(3.36) has a simple relation with the full estimation of the Higgs
polarisation tensor $\Pi_R(Q)$ imaginary part, by means of the effective
perturbation theory. Explicitly, $${\rm Im}{}^*\Pi_R(q,{\vec 0})\simeq \
{\rm Im}\Pi^{{\rm Born}}_R(q,{\vec 0})\ +\ 2\times (3.31)\ +\
(3.36)\eqno(3.37)$$ where, within standard notations, the quantity ${\rm
Im}{}^*\Pi$ corresponds to the picturial representation of Fig.2, and
where ${\rm Im}\Pi^{{\rm Born}}$ is just read from (3.31) or (3.36) by
dropping the $Z/\varepsilon$ and $F$ functions. Once completed with the
results of the next section, the expressions (3.31) and (3.37) will
provide the basis of a future numerical analysis [18].

\bigskip\bigskip\bigskip

{\bf 4- The full HTL treatment} \bigskip Our previous treatment of self
energy insertions is based on (3.17), that is on the leading, singular
most terms in an $\varepsilon$ expansion of the functions $A$ and $B$.
However, nothing guarantees that the less singular parts could be safely
ignored. This possibility even appears somewhat paradoxical. Singular
pieces, generated by this order $\varepsilon^{-1}$-part, usually cancel
out in the calculation of transition rates, as we have just seen, and
are accordingly not expected to be the most relevant ones. Indeed, we
are carrying out the calculation a different way, and we must now show
how our previous results get modified when proper account is taken of
the subleading $\varepsilon$-expansion terms. For the sake of simplicity
this is illustrated by considering zero self energy insertion along the
$P'$ line. \medskip\noindent As a shorthand notation, we introduce $v$,
the variable $$v(\varepsilon) ={1+2^{1+ \varepsilon}\varepsilon \over
4^\varepsilon}\ ,\ \ \ {\rm or}\ \ \  v=1+2\varepsilon\ \ \ \
\longrightarrow\ \ \ \ v=1+\kappa\varepsilon + O(\varepsilon^2)  
\eqno(4.1)$$The
second possibility for $v$ corresponds to the case where factors of
$2^{\pm\varepsilon}$ are dropped, as we have seen in the previous
section 3, whereas the last expression summarizes any of the two
possibilities. The real part of the full HTL self energy reads
accordingly $${\cal A}(P^2,p^2)=(-{m^2\over \varepsilon
p^2}P^2)\left(1-v\vert {P^2\over p^2}\vert^\varepsilon
\left(\Theta(P^2)+\cos (\pi\varepsilon)\Theta(-P^2) \right)
\right)\eqno(4.2)$$ whereas the imaginary part, the function
$B(P^2,p^2)$ of (2.5), can be written $${\cal
B}(P^2,p^2)=-u(\varepsilon)(-P^2)^{1+\varepsilon}\Theta(-P^2) \ ,\ \ \ \
\ \ \ \ \ \ u(\varepsilon)={\pi m^2\over p^{2+2\varepsilon}}\
v(\varepsilon)\ {\sin(\pi\varepsilon)\over \pi\varepsilon} \eqno(4.3)$$
Acting upon ${\cal A}^N$, the terms involving $\delta^{(N)}$
distributions are left the same as before, whereas the only changes come
from the ${\rm Im}\{\Sigma^N\}$-pieces which now read $${\rm
Im}\{\Sigma^N\}=\varepsilon (p_0)\ \sum_{k=0}^{2k=N-1-\Theta((-1)^N)}
(-1)^k C^{2k+1}_N {\cal A}^{N-2k-1}{\cal B}^{2k+1}\eqno(4.4)$$ and where
the sum over $k$ is extended to $k=0$ so as to cover the new singular
and regular perturbative contributions induced by the new functions
${\cal A}$ and ${\cal B}$. Summing over $N$, and taking the large $N$
limit, one has to control the extra pieces which are generated when
$\varepsilon$ is eventually sent to zero. Once folded with the
corresponding principal value distribution of (3.3), the building blocks
of (4.4) yield $$ {\bf \sum}_{ N,k}\ (-1)^k\left( u(\varepsilon)
\right)^{2k+1}\ {1\over \varepsilon}\ \left( C^{2k+1}_{ N}\sum_{j=0}^{ {
N}-2k-1}\ {(-1)^jv^j(\varepsilon)\cos^j(\pi\varepsilon)\over j+2k+1} C_{
{ N}-2k-1}^j\right)\ (-{m^2\over \varepsilon p^2})^{{
N}-2k-1}\eqno(4.5)$$ where a sign $\varepsilon(p_0)$ is understood, as
well as the range $\{1\leq N\leq \infty\}$. Let ${\widehat
{\Pi}}(\cos(\pi\varepsilon))$ be that very expression between
parenthesis in (4.5), and first calculate ${\widehat
{\Pi}}(\cos(\pi\varepsilon))$ at $\cos(\pi\varepsilon)=1$, that is,
${\widehat {\Pi}}(1)$. With respect to the previous calculation,
equations (3.21) to (3.25), the whole change is thus translated into the
occurence of the new factors $u(\varepsilon)$, as defined in (4.3), and
$\{v^j\cos^j(\pi\varepsilon)\}$ between the parenthesis. Setting
$\cos(\pi\varepsilon)=1$, the corrections to (3.21)-(3.25), can be
obtained by Taylor expanding ${\widehat {\Pi}}(1)$ in powers of
$\varepsilon$, writing $${\widehat {\Pi}}(1)=\sum_{n=0}^\infty\
{(\kappa\varepsilon)^n\over n!}{\widehat {\Pi}}^{(n)}(1),\ \ \ \
{\widehat {\Pi}}^{(0)}(1)={1\over 2k+1}\eqno(4.6)$$where the value of
${\widehat {\Pi}}^{(0)}(1)$ is given by (3.24). Unfortunately, when
doing so, the coefficients ${\widehat {\Pi}}^{(n)}$ result in finite
series of Stirling numbers of the first kind. While instructive for some
respects (next section 6 and appendix), the resummation of the ensuing
series is practically undoable [23], but for a few (three) Stirling
numbers. This suggests to follow a different approach. Indeed, a first
order inhomogeneous differential equation in the variable $v$ is readily
found for ${\widehat {\Pi}}(1)$, that we hereafter define as $\Pi (v)$.
Taking the appropriate ``boundary condition" into account, the solution
reads $$\Pi(v)={1\over v^{2k+1}}\ \left(\Pi(1)+C^{2k+1}_N\int_{1}^v{\rm
d}x\ x^{2k}(1-x)^{N-2k-1}\right)\ ,\ \ \ \ \Pi(1)={1\over
2k+1}\eqno(4.7)$$where the value of $\Pi(1)$, the ``boundary condition"
is, again, fixed by (3.24). It is clear that the first term of (4.7),
proportional to $\Pi(1)$, just reproduces a result identical to (3.31),
that is $$\left( {Z\over \varepsilon}={p^2\over m^2}\right)\ {\rm
{tg}}^{-1}\left({u(\varepsilon)\over v(\varepsilon)}\right)\eqno(4.8)$$
whereas the second term of (4.7) yields $$ {\bf \sum}_{ N,k}\
(-1)^k\left( {u(\varepsilon)\over v(\varepsilon)} \right)^{2k+1}\
{1\over \varepsilon}\ \left( C^{2k+1}_{ N}\int_1^v{\rm d}x\ x^{2k}(x
-1)^{N-2k-1}\right)\ (+{m^2\over \varepsilon p^2})^{{
N}-2k-1}\eqno(4.9)$$ Then, relying on an average value theorem for the
integral over $x$, we know that there exists some number
$c(\varepsilon)$ such that (4.9) reads $${1\over c(\varepsilon)} {\bf
\sum}_{ N,k}\ (-1)^k\left( {u(\varepsilon) c(\varepsilon)\over
v(\varepsilon)}  \right)^{2k+1} ({p^2\over m^2}) \ C^{2k+1}_{ N}\
{[{m^2\left(v(\varepsilon)-1\right)/ \varepsilon p^2}]^{N-2k}\over
N-2k}\ ,\ \ \ \ 1\leq c(\varepsilon)\leq v(\varepsilon) \eqno(4.10)$$To
proceed further, it is possible to take advantage of the trick
introduced in the appendix, equation (A.1), and one finds $${1\over
c(\varepsilon)}\ {p^2\over m^2}\Biggl\lbrace {\rm tg}^{-1}\left(
{u(\varepsilon) c(\varepsilon)\over v(\varepsilon)}\left(1
-{m^2(v-1)\over \varepsilon p^2}\right)^{-1} \right) - {\rm
tg}^{-1}\left( {u(\varepsilon)c(\varepsilon)\over
v(\varepsilon)}\right)\Biggr\rbrace \eqno(4.11)$$In order to get (4.11)
out of (4.10), series have been summed up, which are convergent within a
radius specified by the condition $$p^4/ m^4\geq \pi^2+2^2\eqno(4.12)$$
The limit $\varepsilon=0$ can be taken safely. One has $v(0)=1$, and
likewise, $c(0)=1$ by virtue of (4.10). Adding up (4.8) and (4.11), the
full result reads $$\left( {p^2\over m^2}\right)\ {\rm
tg}^{-1}\left({\pi m^2 \over p^2-\kappa
m^2}\right)\eqno(4.13)$$\smallskip\noindent In the calculation of ${\rm
Im}\Pi_R(Q)$, $p$ is fixed at $q/2$ by the kinematics and so, including
the Born term as a constant that we do not write for short, one gets
eventually $$\{C^{st}\}^{-1}{\rm Im}\Pi_R(Q)=({q^2\over 4 m^2})\ {\rm
tg}^{-1}\left({4\pi m^2 \over q^2-4\kappa
m^2}\right)\eqno(4.14)$$\medskip\smallskip\noindent Now, one must
envisage the change in the calculations of (4.5) based on ${\widehat
{\Pi}}(\cos(\pi\varepsilon))$ instead of ${\widehat {\Pi}}(1)$. Indeed,
one can show that (4.8), (4.11) and (4.14) are left the same, but for
terms of order $\varepsilon$ at most. In the calculation of (4.5) based
on ${\widehat {\Pi}}\left(\cos(\pi\varepsilon)\right)$ instead of
${\widehat {\Pi}}(1)$, we get, as a first non trivial correction to
$\Pi(1)$, $$\{{-(\pi\varepsilon)^2\over 2!\ \varepsilon}\}{\bf \sum}_{
N,k}\ (-1)^k\left( u(\varepsilon) \right)^{2k+1}\left( C^{2k+1}_{
N}\sum_{j=0}^{ { N}-2k-1}\ {(-1)^j\ jv^j\over j+2k+1} C_{ {
N}-2k-1}^j\right)\  (-{m^2\over \varepsilon p^2})^{{
N}-2k-1}\eqno(4.15)$$ that is, $$\{{-(\pi\varepsilon)^2\over 2!\
\varepsilon}\}\sum_{ N,k}(-1)^k u(\varepsilon)^{2k+1} C^{2k+1}_N
({\kappa m^2\over p^2})^{N-2k-1}-\{{-(\pi\varepsilon)^2\over
2!}\}{p^2\over m^2}[u(\varepsilon){\partial \over \partial
u(\varepsilon)}]\ \Bigl\lbrace (4.8)+(4.11)\Bigr\rbrace\eqno(4.16)$$The
second term of order $\varepsilon^2$ is obviously finite. The first one
can be calculated and is finite too. One gets $$\{-{\pi^2\over
2!}\varepsilon\}\ \left({\pi m^2p^2\over (p^2-\kappa
m^2)^2+\pi^2m^4}-{\pi m^2\over p^2}\right)\  +\
O(\varepsilon^2)\eqno(4.17)$$Then, discriminating between the cases $j$
even or odd (for the expansion of $\cos^j\left(\pi\varepsilon\right)$),
a similar conclusion is reached for the subsequent, higher order
$\varepsilon$-corrections to ${\widehat {\Pi}}(1)$. Alternatively, one
may proceed a more global way, and try to express ${\widehat
{\Pi}}\left(\cos(\pi\varepsilon)\right)$ as a function of ${\widehat
{\Pi}}(1)$ and $\cos(\pi\varepsilon)$. A first order inhomogeneous
differential equation satisfied by ${\widehat
{\Pi}}\left(\cos(\pi\varepsilon)\right)$ can be derived, the solution of
which reads (with the appropriate "boundary condition", that is
${\widehat {\Pi}}(1)=\Pi(v)$), $${\widehat
{\Pi}}(\cos(\pi\varepsilon))={1\over \cos^{2k+1}(\pi\varepsilon)}\
\left(\Pi(v)+{C^{2k+1}_N\over
v^{2k+1}}\int_v^{v\cos(\pi\varepsilon)}{\rm d}x\
x^{2k}(1-x)^{N-2k-1}\right)\eqno(4.18)$$Using (4.18), it is
straightforward to check that the same result as (4.17) is obtained.

\bigskip\bigskip\bigskip {\bf 5- The mixing of singularities} \bigskip A
well known feature of real time formalisms is the fact that $T=0$ and
$T\neq 0$ parts come out additive. This property of first non trivial
order of ordinary perturbation theory does not persist at higher orders
which mix up zero and non-zero temperature contributions. When the
latter are singular, the situation is the one of "singularity mixing",
an admittedly complicated one [9,12], which, at least at the author
knowledge, has not been much investigated until now. In the present
situation though, the singularity mixing problem turns out to be
unexpectedly simplified and this is worth emphasizing.\medskip\noindent
At $T=0$, the one loop order self energy is well known [9,13]
$$\Sigma(P^2,\mu^2)={\alpha\over 2}B({D\over 2}-1, {D\over 2}-1 )\
P^2(-{P^2\over \mu^2})^{{D\over 2}-3}\ ,\ \ \ \ \ \alpha={g^2\over
(4\pi)^{D/2}} \eqno(5.1)$$Renormalized at mass scale $\mu$, one
therefore has, writing $\Sigma_F =a+ib$, $$a(P^2,\mu^2)=-{\alpha \over
12\varepsilon}P^2\left(1-\vert{P^2\over \mu^2}
\vert^\varepsilon\left(\Theta(-P^2)+\Theta(P^2)\cos(\pi\varepsilon)\right)
\right),\ \ \ \ \ b(P^2)=-{\pi\alpha \over
12}\left(P^2\right)^{1+\varepsilon}\Theta(P^2){\sin(\pi\varepsilon)\over
\pi\varepsilon } \eqno(5.2)$$In the function $a(P^2, \mu^2)$, the first
term, given by the $+1$ in the parenthesis, is the ultraviolet
counter-term of the $T=0$ renormalization procedure. When differentiated
with respect to $P^2$, though, the same factor $1/\varepsilon$ is to be
understood as the dimensional regularization of an infrared (collinear)
singularity, as is made obvious by simple power counting
arguments.\medskip\noindent Now, a $T\neq 0$ real time formalism can be
used to calculate pure $T=0$ quantities, just by letting $T$ tend to
zero. This long known property of real time formalisms [14] has been
explicitly verified in [9] at second non trivial order of perturbation
theory. Thus, in principle, the global one loop self energy correction
can be inserted in our previous equations, so as to get by the same
token, both zero and non zero temperature corrections. In practice
however, the different analytical properties involved at $T=0$ and
$T\neq 0$, have long made this property short of any practical
purpose.\smallskip\noindent Here instead, we will take advantage of the
results of previous sections 3 and 4, dealing first with the functions
$A$ and $B$ of section 3. Then, remarking that $D_u$, the infinitesimal
dilatation operator in the variable $u_{\varepsilon=0}(p)$, ignores both
$Z(\varepsilon,p)/\varepsilon$ and $T=0$ variables, the results are made
complete by acting upon them with $D_u$, $${\rm
exp}\left(-\ln\left(1-{m^2(v(\varepsilon)-1)\over \varepsilon
p^2}\right)\ D_u\right)\ {\rm tg}^{-1}\left(u\right)={\rm
tg}^{-1}\left(u\left( 1-{\kappa m^2\over p^2}\right)^{-1}\right)\ ,\ \ \
D_u=u{\partial \over \partial u}\eqno(5.3)$$

\smallskip\noindent Thus, one can re-write those relevant parts of (2.3)
and (2.5) the following way $$A(P^2,p^2)=T^{2\varepsilon}\ ({2\pi T\over
p})^2 \left(-{\alpha \over 12\varepsilon}P^2\left(1-\vert{P^2\over p^2}
\vert^\varepsilon (\Theta(P^2)+\Theta(-P^2)\cos(\pi\varepsilon))
\right)\right)\eqno(5.4)$$and $$B(P^2,p^2)=({T\over p})^{2 \varepsilon}\
({2\pi T\over p})^2\left(-{\pi \alpha \over 12}
(-P^2)^{1+\varepsilon}\Theta(-P^2){\sin(\pi\varepsilon)\over
\pi\varepsilon } \right)\eqno(5.5)$$A global one loop renormalized
(Feynman) self energy can accordingly be defined as $${\widehat
\Sigma}_F(P^2,p^2,\mu^2)={\widehat A}(P^2,p^2,\mu^2)+i{\widehat
B}(P^2,p^2)\eqno(5.6)$$ with $${\widehat
A}(P^2,p^2,\mu^2)=a(P^2,\mu^2)-({2\pi T\over p})^2\
a(-P^2,p^2)\eqno(5.7)$$ $${\widehat B}(P^2,p^2)=b(P^2)+({2\pi T\over
p})^2\ b(-P^2)\eqno(5.8)$$ Folded with the principal value
distributions, over their respective kinematical domains, the functions
$b(P^2)$ and $b(-P^2)$ involve the same pole structure, and, up to terms
of order $(p^2/ \mu^2)^\varepsilon $, the associated residues are the
same. In these topologies however, these terms do not regularise any
singular behaviour and their $\varepsilon=0$ limit can be taken with the
alluded result (likewise, for the same reason, writing (5.7) and (5.8),
the $ T^{2\varepsilon}$ and $(T/ p)^{2 \varepsilon}$ terms of (5.3) and
(5.4) have been ignored).

\medskip\noindent As a consequence, the whole series we have just
analysed in the pure thermal case, goes through, unchanged, in the
global treatment of both $T=0$ and $T\neq 0$ parts, but for the two
replacements $${m^2\over \varepsilon p^2}\Longrightarrow {\alpha\over
12\varepsilon}\left(1-\left(-({2\pi T\over p})^2\right)\right)$$ $$
u_{\varepsilon=0}(p)={\pi m^2\over p^{2}}\Longrightarrow {\pi m^2\over
p^{2}}(1-{m^2(v(\varepsilon)-1)\over \varepsilon
p^2})^{-1}\Longrightarrow {\pi\alpha\over 12}\left(1+({2\pi T\over
p})^2(1-{m^2(v(\varepsilon)-1)\over \varepsilon
p^2})^{-1}\right)\eqno(5.9)$$ Result (4.14), for example, is transformed
into $$\{C^{st}\}^{-1}{\rm Im}{\widehat \Pi}_R(Q)= \left({\widehat
Z}(\varepsilon,q/2)\over \varepsilon\right) {\widehat
F}(q/2)\eqno(5.10)$$ where we have defined the global functions (adding
up the Born term) $${\widehat F}(p)={\rm tg}^{-1}\Biggl\lbrace
{\pi\alpha\over 12}\left(1+({2\pi T\over p})^2\
(1-{m^2(v(\varepsilon)-1)\over \varepsilon p^2})^{-1}
\right)\Biggr\rbrace\eqno(5.11)$$ $${\widehat Z}^{-1}
(\varepsilon,p)=1+{\alpha\over 12\varepsilon }\left(1+({2\pi T\over
p})^2\right)\eqno(5.11)$$ Expressions (5.11) and (5.12) are obviously
reminiscent of the elementary first order additivity of the $T=0$ and
$T\neq 0$ parts, while displaying its full one loop leading order (HTL)
realization. Of course, we do not expect that this simple pattern should
extend beyond the leading HTL approximation. Clearly, the HTL
peculiarity is at play (see remark (i) below). By construction, the
thermal contributions of (5.11) and (5.12) are valid for soft internal
momenta and are thus enhanced by a factor $1/g^{2}$ with respect to the
$T=0$ ones. Some remarks are in order.\medskip (i)-The striking
similarity between the renormalized $T=0$ self energy and its $T\neq 0$,
HTL counter part is worth emphasizing (compare equations (5.2) with
(5.4) and (5.5) ). It is at the origin of a possible global treatment,
as we have just sketched. Unfortunately, we have been unable to find any
simple interpretation of this amazing fact, except that it is certainly
in line with [21]. \smallskip (ii)-Things can be viewed the other way
round. Once it is verified that the couples of functions $(a,b)$ and
$(A,B)$ are involved in phase space domains over which they develop the
same singular structures, then, (5.7) and (5.8) obviously indicate that
the issues of singularity compensations will be the same in either
cases. For the topologies considered here, where only mass singularities
show up, the above statement is nothing but an alternative form of a
Niegawa's recent result [15]. \bigskip\bigskip\bigskip {\bf 6.
Renormalization constants}\bigskip In the past twelve years, efforts
have been sometimes devoted to the possibility of defining
renormalization constants in a thermal context [16]. When fermionic
fields are involved, for example, such a notion could be shown to be
almost unreliable [16], whereas more room was left in the case of
Bose-Einstein statistical fields [17]. However, even in the simpler
scalar case, some ambiguity was left regarding its definition.
\medskip\noindent According to [17], the external leg renormalisation
constant should be defined as $$Z(p^2)={ \rm Re}\left(1- \Sigma'
(0,p^2)\right)^{-1}\eqno(6.1)$$where the $P^2$-derivative indicated by
the prime must be identified with a total derivative, kinematics
rendering $p^2$ a function of $P^2$, process dependent, though weakly.
At $T=0$, partial and total derivatives coincide because of Lorentz
invariance, and the definition of $Z$ reads instead $$Z=\left(1-{ \rm
Re}\Sigma'(0)\right)^{-1}\eqno(6.2)$$ Besides the familiar lack of
Lorentz invariance inherent to non zero temperatures, and reflected in
the $p$-momentum dependence, definition (6.1) differs significantly from
(6.2) by the very place where the real part prescription, ${\rm Re}$,
should be taken. Furthermore, it is worth remarking that, if pertinent,
(6.1) should also apply at $T=0$ for unstable particles [17]. In a few
particular cases both definitions agree. For massless fields within the
dimensional regularization scheme (see (2.3) and (2.5)), for example,
and also in the case of a real valued self energy function
$\Sigma(P^2,p^2)$, because of the $+i\varepsilon_\alpha p_0$ advanced or
retarded prescription which completes the dressed propagator
determination. But the one loop self energy is not real valued in
general.\medskip\noindent Indeed, the identification (6.1) is obtained
by a formal resummation of the terms involving
$\delta^{(N)}(P^2)$-distributions only, as they appear on the right hand
side of (3.3). Now, our analysis has shown that these contributions can
not be disentangled from those involving principal value distributions,
folded with imaginary parts of self energy insertions. When the latter
are non zero, that is, when the self energy is complex valued, it was
recognized in [17] that an expansion in the number of self energy
insertions was necessary in order to give a well defined meaning to the
calculations. \smallskip\noindent The analysis of the previous sections
relies on such an expansion and is accordingly able to identify
$$Z(p^2)=\left(1- {\rm Re}\Sigma'(0,p^2)\right)^{-1}\eqno(6.3)$$ rather
than (6.1), as the correct renormalisation constant, thermal counterpart
(see the appendix). \bigskip\bigskip\bigskip\bigskip\bigskip\bigskip
{\bf Appendix} \bigskip \medskip \medskip \bigskip {\bf {On section
3}}\bigskip As the arithmetical identity (3.24) is used extensively
throughout the article, we find worthwhile to give its demonstration,
the more as a proof by induction appears hopeless. Rather we can write
$$(3.24)=C_{{ N}}^{2k+1}(\sum_{j=0}^{{ N}-2k-1}\ (-1)^j C_{{\cal
N}-2k-1}^j)\int_0^1 {\rm d}x\ x^{j+2k}\eqno(A.1)$$Now, interchanging the
sum with the integral, one obtains the alluded identity as $$C_{{
N}}^{2k+1}\int_0^1 {\rm d}x\ x^{2k}(1-x)^{N-2k-1}=C_{{
N}}^{2k+1}B(2k+1,N-2k)={1\over 2k+1}\eqno(A.2)$$ where $B(x,y)$ is the
Euler beta-function. A whole series of similar arithmetical identities
can be obtained that way.
\bigskip\medskip\bigskip\medskip\bigskip\medskip\bigskip\medskip

Here are given the basic but rather cumbersome expressions leading from
(3.32) to (3.36) . From (3.32), let us write explicitly $${\rm
Im}\left(\Delta^{(N')}_R(P')\right)={\bf P}{{\rm
Im}\Sigma_R^{N'}(P')\over (P'^2)^{N'+1}} -\pi{(-1)^{N'}\over
N'!}\varepsilon(p'_0)\delta^{(N')}(P'^2){\rm
Re}\Sigma_R^{N'}(P')\eqno(A3)$$Consider ${\rm Im}\Pi^{(N,N')}_R(Q)$,
with $N'$ fixed, and write the contribution of a generic term appearing
in ${\rm Im}\Sigma^{N'}_R(P')$. One gets $$\displaylines{\int {{\rm
d}^{D-1}p\over 2p_0(P^2,p^2)}(1+2n^B(p_0))\sum_{k'=0}^{2k'
=N'-1-\Theta((-1)^{N'})}C^{2k'+1}_{N'}(-1)^{k'}\varepsilon(p'_0)(-{\pi
m^2\over p'^{2+2\varepsilon}})^{2k'+1}(-{m^2\over \varepsilon
p'^2})^{N'-2k'-1}\cr\hfill\int {\rm d}P^2 {\Theta (-P'^2){\cal {\bf
P}}\over (-P'^2)^{1-(2k'+1)\varepsilon}}(1-\vert {P'^2\over
4p'^2}\vert^\varepsilon)^{N'-2k'-1}\biggl\lbrace
-\pi\delta(P^2)(-{m^2\over \varepsilon p^2})^N  -{{\cal {\bf P}}\over
(P^2)^{N+1}}\cr\hfill \sum_{k=0}^{2k=N-1-\Theta((-1)^N)}C^{2k+1}_N
(-{m^2\over \varepsilon p^2})^{N-2k-1}\left(P^2(1-\vert {P^2\over
4p^2}\vert^\varepsilon)\right)^{N-2k-1}\cr\hfill(-1)^k(-{\pi m^2\over
p^{2+2\varepsilon}})^{2k
+1}((-P^2)^{1+\varepsilon})^{2k+1}\Theta(-P^2)\biggr\rbrace\qquad(A4)}$$In
the light cone neighbourhood, the sum over $P^2$ develops mass  
singular
behaviours , whereas the remainder of the integrand is a well behaved
function. The singular part of $(A4)$ is accordingly given by
$$\displaylines{\int {{\rm d}^{D-1}p\over
2p}(1+2n^B(p))C^{2k'+1}_{N'}(-1)^{k'}\varepsilon(q-p)(-{\pi m^2\over
p^{2+2\varepsilon}})^{2k'+1}(-{m^2\over \varepsilon
p^2})^{N'-2k'-1}\cr\hfill \Theta (2p-q)[q(2p-q)]^{-1+(2k'+1)\varepsilon}
\left(1-\vert {q(2p-q)\over 4p^2}\vert^\varepsilon\right)^{N'-2k'-1}\int
{\rm d}P^2\biggl\lbrace -\pi\delta(P^2)(-{m^2\over \varepsilon p^2})^N
\cr\hfill -{{\cal {\bf P}}\over (P^2)^{N+1}} C^{2k+1}_N (-{m^2\over
\varepsilon p^2})^{N-2k-1}\left(P^2(1-\vert {P^2\over
4p^2}\vert^\varepsilon)\right)^{N-2k-1}\cr\hfill(-1)^k(-{\pi m^2\over
p^{2+2\varepsilon}})^{2k
+1}((-P^2)^{1+\varepsilon})^{2k+1}\Theta(-P^2)\biggr\rbrace\qquad(A5
)}$$ where we have used the relation ${\vec p}={\vec p'}$, particular to
both the physical process under consideration with $Q=(q,{\vec 0})$, and
our notations, specified on Fig.1. Integrating now over $P^2$, the
singular most terms of order $\varepsilon^{-N}$ cancel out thanks to
(3.20), leaving a finite series of sub-leading mass singularities. The
latter can be summed over $N$, using (3.24 ), so as to yield eventually
$$\displaylines{\sum_N^\infty (A2)=\int {{\rm d}^{D-1}p\over
2p}(1+2n^B(p))\sum_{k'=0}^{2k'=N'-1-\Theta ((-1)^{N'})}
C^{2k'+1}_{N'}(-1)^{k'}\varepsilon(q-p)(-{\pi m^2\over
p^{2+2\varepsilon}})^{2k'+1}\cr\hfill(-{m^2\over \varepsilon
p^2})^{N'-2k'-1} {\Theta (2p-q)\over (q(2p-q))^{-1+(2k'+1)\varepsilon}}
(1-\vert {q(2p-q)\over
4p^2}\vert^\varepsilon)^{N'-2k'-1}({Z(\varepsilon,p)\over \varepsilon})
F(p)\qquad(A6)}$$The integral over ${\vec p}$ factorizes the total solid
angle of the model. Then, the integration over $|{\vec p}|$ ranges from
$p=q/2$ to $p=\infty$ because of the $\Theta (2p-q)$ distribution, but
the sign distribution $\varepsilon (q-p)$ splits the integral into two
parts. The second part, say, with an integration range from $p=q$ to
$p=\infty$, will be discarded because, then, one is leaving the domain
of softness of the $P,P'$ lines, that is, the realm of relevance of the
effective perturbation theory. Even though, it is easy to check that
this neglected integral is perfectly regular. The first part, with an
integration range from $p=q/2$ to $p=q$, exhibits light cone singular
behaviours at the $p=q/2$ lower bound, through the factors (3.33). Now,
by using (3.35), it is straightforward to show that the singular most
contribution, of order $\varepsilon^{-N'}\times (q^2/4m^2)\times F(q/2)$
coming from the term $k'=0$ in (A6), cancels out against the
contribution coming from the ${\rm Re}\Sigma_R^{N'}(P')$ term of $(A3)$
exactly like for the $P$ line $$ -\pi{(-1)^{N'}\over
N'!}\varepsilon(p'_0)\delta^{(N')}(P'^2){\rm
Re}\Sigma_R^{(N')}(P')\Longrightarrow -\pi\varepsilon(p'_0)\delta(P'^2)
\left(-{m^2\over \varepsilon p'^2}\right)^{N'}\eqno(A7)$$ In (A6), one
is left with the sum of $\{B^{2k'+1}\}$-terms, at $k'\geq 1$, which
entail the factor $\{\left(q(2p-q)\right)^{-1+(2k'+1)\varepsilon}\}$,
singular at the lower boundary of the $p$-integration, whereas the
remainder of the integrand, that is, basically, the functions
$Z(\varepsilon,p)/\varepsilon$ and $F(p)$, is perfectly regular. Then it
is easy to use the arithmetical identity (3.24) in order to show that
the same pattern as for the $P$-line applies here again, with a finite
series of sub-leading mass singularity outcome. Summing the latter over
$N'$, equation (3.36) results in the limit $N'=\infty$.
\bigskip\medskip\bigskip\medskip\bigskip\medskip\bigskip\medskip

{\bf {On section 6}}\bigskip In section 3, equation (3.26), we have
pointed out the occurence of a global factorizing expression
$Z(\varepsilon, p)$. Together with the overall factor $1/\varepsilon$,
the function $Z$ contributes the global simple function $p^2/m^2$ (as
seen from (3.29)), which is recovered also throughout the analysis
performed insection 4. In this latter section though, the function $Z$
is not itself immediately apparent, thereby obscuring somehow its
interpretation as a renormalization constant. It is indeed an artefact
of the approach followed in section 4.  This is realized by getting back
to the $\varepsilon$-expansion (4.6), for which we have ($N-2k-1\geq n$)
$${\widehat \Pi}^{(n)}(1)= C^{2k+1}_{ N}\sum_{j=0}^{ { N}-2k-1}\
{(-1)^jj(j-1)..(j-n+1)\over j+2k+1} C_{ { N}-2k-1}^j \eqno(A.9)$$Now, by
an iterated use of of the arithmetical identity (3.24), and also of the
identity [11] $$\sum_{j=0}^{ { N}-2k-1}\ (-1)^j\ j^{\ {i-1}}\ C^j_{{
N}-2k-1}=0,\ \ \ \ \ 1\ \ \leq i\leq { N}-2k-1 \eqno(A.10)$$it is easy
to show that any of the ${\widehat \Pi}^{(n)}(1)$ is a polynomial of
degree $(n-1)$ in the variable $(2k+1)$. Explicitly $${\widehat
\Pi}^{(n)}(1)= \sum_{m=0}^{n-1}(-1)^mS^m_{n-1}(2k+1)^m\ ,\ \ \ \ n\geq 1
\eqno(A.11)$$ where the coefficients $\{S^m_{n-1}\}$ are integers,
namely, Stirling numbers. This can be realized by quoting the relation,
due to (A.10), $${\widehat \Pi}^{(n)}(1)=-[(n-1)+(2k+1)]\ {\widehat
\Pi}^{(n-1)}(1),\ \ \ \ \ n\geq 1 \eqno(A.12)$$ and this yields
$${\widehat \Pi}^{(n)}(1)=(-1)^n\biggl\lbrace\prod_{m=0}^{n-1}\left(m+
[2k+1]\right)\biggr\rbrace \ \left({1\over 2k+1}\right) \eqno(A.13)$$
where the generating relation for the Stirling numbers of the first kind
is manifest [23].\smallskip\noindent The point is that at any given
order $\varepsilon^n$ in the expansion (4.6), the coefficient ${\widehat
\Pi}^{(n)}(1)$ is independent of ${ N}$. On view of (4.5), we conclude
that the summation over ${ N}$ can be carried out, and that the ensuing
expression $\{Z(\varepsilon, p)/\varepsilon\}$ keeps entering the
result, for all $n$ and all $k$, as an overall multiplicative factor.

\vfill\eject

{\bf 7. Conclusion} \bigskip On the basis of a scalar model of field
theory, renormalizable and asymptotically free, the light cone singular
behaviour of the corresponding thermal effective perturbation theory has
been investigated for some given process.\medskip\noindent Using the HTL
form of the self energy near the light cone where mass singularities
develop, one can conclude that, at any number of self energy insertions,
a full compensation of singularities is obtained. This conclusion brings
a correction to the erroneous statement which appeared in [8], and was
due to an insufficient dimensional expansion of the self energy
functions. In this respect, it may be recalled that after a first series
of calculations performed at first and second non trivial orders of
thermal perturbation theory [3,9,19], it has been observed (and then
conjectured) that singularity cancellation took place within each
topology separately [20]. We can remark that our analysis supports and
enforces this conjecture.\medskip\noindent Once the detailed balance
compensation of mass singularities is achieved, the result for the
process under consideration, can be obtained by summing over $\{N,N'\}$
the regular contributions which are left, as this summation is mandatory
in order to get the full leading order correction $O(g^2T^2)$ to the
Born approximation.\smallskip \medskip\noindent Throughout the present
analysis, however, the summation has been carried out a somewhat
different way. Although the full HTL self energy is recognized a
perfectly regular function of $P^2$ in any light cone neighbourhood, we
have stressed that it entails a potentially singular most part, dealt
with in section 3, and which turns out to yield the basic result.
Remarkably enough, at the level of approximation where we have been
working, this basic result enjoys two interesting properties. First, it
is immediately related to the cases in which the HTL self energies are
given by the (second kind) Legendre function $Q_0$ (fermionic fields)
and is therefore relevant to the study of thermal QED and QCD, the
infrared problem included [22]. Second, it is simply enough related to
the case under consideration, where one has to cope with a HTL self
energy on the order of the Legendre function $Q_1$ rather than $Q_0$.
\smallskip\noindent Furthermore, proceeding that way, a striking analogy
can be put forth between the $T=0$ renormalized self energy and its HTL
counterpart. This is certainly a remarkable and intriguing aspect which,
at least in our opinion, can only be read in line with the ideas
developed in [21]. In particular, the issued mass singularity structures
(poles and residues), turn out to be identical, and a global treatment
of both zero and non zero temperature corrections is made possible,
resulting in a simple resolution of the otherwise intricate singularity
mixing problem. Likewise, a new, unambiguous identification of a
renormalization constant in the thermal context is obtained, and turns
out to be at variance with some previous attempts. \medskip\noindent The
function $F(p)$, and {\it a fortiori}, its generalization ${\widehat
F}(p)$, is new and interesting in this context, as it could not be
obtained in previous related works [12], [18], where the effective
perturbation theory was used right from the onset. Remarking the
similarity of $F$ with a first order brehmsstrahlung function, one
recovers a physical picture of the HTL's which is closely related to
their interpretation in terms of forward Thompson scattering amplitudes
[21].\medskip\noindent Eventually, we can stress that our guiding
strategy has been to express the singular structure of an effective
quantum field theory at temperature $T$, in terms of the same bare
temperature quantum field theory. Such a strategy may be thought of as
arduous. On the other hand, though, the complexity of the effective
perturbation theory, the mixing of topologies it envolves, renders
hazardous the control of such fine tuning mechanisms as singularity
compensations, if not the very meaning of the ensuing finite results.
The same approach is extended to the case of gauge field theories for
timelike and lightlike external momenta [22]. \vfill \eject {\bf
Acknowledgement} \medskip It is a pleasure to thank R. Pisarski and, in
particular, M. Le Bellac for a very careful reading of the manuscript
and many fruitful comments.\smallskip This research is supported in part
by the EEC Programme "Human Capital and Mobility", Network "Physics at
High Energy Colliders", contract CHRX-CT93-0357 (DG 12 COMA) \

\bigskip\bigskip\bigskip\bigskip\bigskip {\bf References}

\bigskip [1] N.P. Landsman, "Quark Matter 90", Nucl.Phys.
A525(1991)397\smallskip [2] E. Braaten and R. Pisarski,
Phys.Rev.Lett.64(1990)1338; Nucl.Phys.B337(1990)569.\smallskip J.
Frenkel and J.C Taylor, Nucl.Phys.B334(1990)199\smallskip [3] Banff/Cap
Workshop on Thermal Field Theory, F.C. Khanna, R. Kobes,\smallskip G.
Kunstatter and H. Umezawa Editors, World Scientific.\smallskip [4] R.
Baier, S. Peign\'e and D. Schiff, Z.Phys.C62(1994)337.\smallskip [5] P.
Aurenche, T. Becherrawy and E. Petitgirard, ENSLAPP A452,93, HEP-PH
93,(unpublished)\smallskip E. Petitgirard, Th\'ese pr\'esent\'ee \`a
l'universit\'e de Savoie, le 25 F\'evrier 1994\smallskip P. Aurenche and
T. Becherrawy, Nucl.Phys.B379(1992)259\smallskip [6] A. Niegawa,
OCU-PHYS 153, to appear in Mod.Phys.Lett.A (1995)\smallskip [7] F.
Flechsig and A.K. Rebhan,
Nucl.Phys.B464(1996)279;hep-ph/9509313\smallskip [8] T. Grandou,
Phys.Lett.B367(1996)229\smallskip [9] T. Grandou, M. Le Bellac and D.
Poizat, Phys.Lett.B249(1990)478; Nucl.Phys.B358(1991)408;\smallskip M.
Le Bellac and P. Reynaud, Nucl.Phys.B380 (1992)423\smallskip [10] P.
Aurenche, E. Petitgirard and T. del Rio Gaztelurrutia,
Phys.Lett.B297(1992)337\smallskip [11] I.S. Gradshteyn and I.M. Ryzhik,
Table of integrals, series and products \smallskip [12] "Recent
developments in finite temperature quantum field theories",\smallskip M.
Le Bellac, Cambridge University Press, 1996\smallskip [13] A.J.
Macfarlane and G. Woo, Nucl.Phys.B77(1974)91; J.C. Collins,
"Renormalization",\smallskip Cambridge Monographs On Mathematical
Physics, Cambridge University Press, 1984 \smallskip [14] N.P. Landsman
and Ch. Van Weert, Phys.Rep.145(1987)142\smallskip [15] A. Niegawa,
Phys.Rev.Lett.71(1993)3055\smallskip [16] J.F. Donoghue and B.R.
Holstein, Phys.Rev.D28(1983)340; J.F. Donoghue and B.R.
Holstein\smallskip and R.W. Robinett,
Ann.Phys.(N.Y.)164(1985)233;\smallskip R. Kobes and G.W. Semenoff,
Nucl.Phys.B272(1986)329; W. Keil, Phys.Rev.D40(1989)1176;\smallskip Y.
Gabellini, T. Grandou and D. Poizat,
Ann.Phys.(N.Y.)202(1990)436;\smallskip R.D. Pisarski,
Phys.Rev.Lett.63(1989)1129; H.A. Weldon, Physica A158(1989)169\smallskip
[17] M. Le Bellac and D. Poizat, Z.Phys.C47(1990)125\smallskip  [18] T.
Grandou and P. Reynaud, work in progress\smallskip [19] M. Le Bellac and
P. Reynaud, Nucl.Phys.B416(1994)801\smallskip [20] T. Altherr and T.
Becherrawy, Nucl.Phys.B330(1990)174\smallskip [21] G. Barton,
Ann.Phys.(N.Y.)200(1990)271; J. Frenkel and J.C Taylor,
Nucl.Phys.B374(1992)156\smallskip  [22] T. Grandou, in
preparation.\smallskip  [23] Handbook of Mathematical Functions,
National Bureau of Standards,\smallskip M Abramowitz and I. Stegun
Editors, 1966\smallskip [24] Miniworkshop on Thermal Field Theories,
Viena, 16-18 March 1996.

\end